\documentclass{iau307}
\usepackage{graphicx}
\usepackage{natbib}
\usepackage{url}
\usepackage{dtklogos}
\bibpunct{(}{)}{;}{a}{}{,}

\title[Asteroseismology of Massive Stars] 
{Asteroseismology of Massive Stars : Some Words of Caution}

\author[A. Noels et al.]   
{A. Noels$^1$
 \and M. Godart$^2$ \and S. J. A. J. Salmon$^1$ \and M. Gabriel$^1$ \and J. Montalb\'an$^1$ \and A. Miglio$^3$}

\affiliation{$^1$Institut d'Astrophysique et de G\'eophysique, Li\`ege University, All\'ee du 6 Ao\^{u}t, 17, B-4000 Li\`ege, Belgium \\ email: {\tt Arlette.Noels@ulg.ac.be} \\[\affilskip]
$^2$Dept. of Astronomy, The University of Tokyo, 7-3-1 Hongo, Bunkyo-ku Tokyo, 113-0033, Japan \\[\affilskip] 
$^3$School of Physics and Astronomy, University of Birmingham, Birmingham, B15 2TT, UK}

\pubyear{2014}
\volume{307} 
\pagerange{}
\jname{New windows on massive stars: asteroseismology, interferometry, and spectropolarimetry}
\editors{G. Meynet, C. Georgy, J.H. Groh \& Ph. Stee, eds.}

\begin{document}

\maketitle

\begin{abstract}
{Although playing a key role in the understanding of the supernova phenomenon, the evolution of massive stars still suffers from uncertainties in their structure, even during their ``quiet" main sequence phase and later on during their subgiant and helium burning phases. What is the extent of the mixed central region? In the local mixing length theory (LMLT) frame, are there structural differences using Schwarzschild or Ledoux convection criterion? Where are located the convective zone boundaries? Are there intermediate convection zones during MS and post-MS phase, and what is their extent and location? We discuss these points and show how asteroseismology could bring some light on these questions.}
\keywords{stars: evolution, stars: interiors, stars: oscillations (including pulsations), stars: variables: massive stars, stars: mass loss, }

\end{abstract}

\firstsection 

\section{Introduction}
The key role played by asteroseismology in probing and understanding the stellar structure of massive stars is undeniable. Thanks to excited modes, in particular those propagating in the deep interior, the extent of the mixed central region can be constrained. This region consists of the fully mixed convective core surrounded by an ``extra-mixed" region, either fully or partially mixed. The physical origin of this region can be rotation, overshooting... What asteroseismology will test is of course not the physics but the resulting chemical composition profile. In Sect. \ref{extra-mixing}, we show how the extent of the extra-mixed region can only be reliably determined by asteroseismology provided some rather strong requirements are fulfilled not only on the number of well identified modes but also on the detailed elemental abundances of the observed star \citep{Salmon14}.

One of the big issues extensively discussed in this symposium is the treatment of convection in massive stars. However, most stellar evolution codes still compute convective regions in the frame of the local mixing length theory (LMLT) with either Ledoux or Schwarzschild criterion. We first show that whatever the adopted criterion, the extent of convective cores must be identical (see Sect. \ref{criterion}).

Asteroseismic analyses require fully consistent models, in particular models with correctly located convective boundaries, \textit{i.e.} satisfying the convection criterion on its convective side. As shown in \citet{Gabriel14}, a departure from this requirement leads to too small convective cores and erroneous Brunt-V\"ais\"al\"a frequency distributions, with a possible misleading of the asteroseismic interpretation. This is discussed in Sect. \ref{boundary}.

In Sect. \ref{shell} we address the important point of the exact location and extent of convective shells in the vicinity of the hydrogen burning shell in supergiant B stars and we discuss the different aspects of these shells resulting from the choice of Ledoux or Schwarzschild convection criterion. We raise some problems (semi-convection, overshooting, non-existence of a static solution) that can appear when a convective shell develops \citep{Gabriel14}. 

We then show in Sect. \ref{BSup} how recent interesting asteroseismic analyses can help bring new constraints on these shells \citep{Saio06, Gautschy09, Godart09, Godart14, Saio13, Georgy14}.

Sect. \ref{opacity} addresses the problem of exciting $\beta$ Cephei- and SPB-type modes in the Small Magellanic Cloud, with a metallicity so low that no such modes (only very few SPB-type) are indeed theoretically expected. This could be due to an underestimation of the Ni opacity in the ``metal opacity bump" \citep{Salmon12}. 

\section{Amount of extra-mixing in MS and core helium burning stars}
\label{extra-mixing}
In what follows we shall define the extra-mixing as an additional full mixing taking place on top of the convective core, resulting from various physical causes such as convective overshooting or penetration, rotation, semi-convection, diffusion... The extent of this extra-mixing region is referred to as a fraction $\alpha_{em}$ of the pressure scale height. 

As was shown and discussed in session II of this symposium \citep{Aerts14}, several asteroseismic analyses of $\beta$ Cephei stars have revealed quite a large range of values for this parameter $\alpha_{em}$, from 0 to $\sim$ 0.5. The method used is a $\chi^2$ minimisation of the differences between the observed frequencies and those found in a large grid of models with various stellar parameters, in particular various values of $\alpha_{em}$. No direct relation between this latter parameter and another stellar property, such as the rotation velocity, seems to exist. 

In an attempt to better understand the link between $\alpha_{em}$ and the fitting method, \citet{Salmon14} has realized a series of \textit{hare-and-hound} exercises. Targets were computed with various choices of stellar parameters (M, R, X, Z, $\alpha_{em}$) and different physical assumptions, such as the solar mixtures AGSS05 \citep{Asplund05} or GN93 \citep{Grevesse93} and the opacities OPAL \citep{Iglesias96} or OP \citep{Badnell05}. A set of non-adiabatic frequencies were computed for each target, to serve as ``observed" frequencies.The grid of models used to minimize $\chi^2$ for each target had fixed physical assumptions, \textit{i.e.} AGSS05 for the solar mixture and OP for the opacities. Adiabatic frequencies were available for each model of the grid. We cite below some of the conclusions drawn from these exercises~:

\begin{itemize}
\item If the physics is the same in the target and in the grid, five \textit{well identified} (well known) observed frequencies are required to recover the stellar parameters, in particular $\alpha_{em}$. With only three frequencies, the error on $\alpha_{em}$ can reach a factor two.
\item If among those five frequencies, one of them is not correctly identified, the errors are as large as in the case where only three frequencies are observed. The minimum of $\chi^2$ is however much larger than when a good solution is obtained.
\item If the target is built with the solar mixture GN93 instead of that of the grid (AGSS05), a 50 \% error is found on $\alpha_{em}$. Here also the minimum of $\chi^2$ presents a rather large value that could be used as an alarm criterion to detect an incorrect solution. 
\end{itemize}

It should however be noted that, without asteroseismic data, the amount of extra-mixing in massive stars is impossible to obtain since there is a well-known degeneracy in $M$ and $\alpha_{em}$ for a given set of $T_{\textrm{eff}}$ and $L$. But we think useful to stress the importance of combining asteroseismic data with spectroscopic and photometric data in order to reach a higher level of constraints on the model. Moreover, ``best" solutions obtained with a too large $\chi^2$ should be regarded with caution.

\section{Convection in LMLT frame}
\label{convection}
In this section, we would like to emphasize, and call attention on, some problems related to convective zones and, in particular to convective boundaries, which seem to be met in some stellar evolution codes. Since some codes, like the MESA code for instance \citep{Paxton13}, are now being used on a larger and larger scale by the astrophysical community, extra care should be given to the outputs in order to help the developers to still increase the performances of the code. Another reason to be extremely careful is that an asteroseismic analysis requires a fully consistent model, at the risk of misinterpreting ateroseismic data if this exigence is not fully met.
 
\subsection{Convection criterions}
\label{criterion}
In spherically symmetric stars, the condition to define the boundary of a convective zone is \citep[see][and references therein]{Gabriel14}
\begin{equation}
V_r = 0
\end{equation}
where $V_r$ is the radial component of the convection velocity. In the frame of the LMLT, this implies
\begin{equation}
L_{rad} = L(m) \;\; \mathrm{and} \;\; \nabla_{rad} = \nabla_{ad} \label{Sch}
\end{equation}
where $L_{rad}$ is the radiative luminosity and $L(m)$, the total luminosity at mass fraction $m$; $\nabla$ stands for usual temperature gradient and the indices $rad$ and $ad$ refer to radiative and adiabatic. This is the so-called Schwarzschild criterion. Let us visualize a convective boundary as a two-sided spherical surface, with a ``convective" and a ``radiative" side. The condition \ref{Sch} is obviously only meaningful in a convective region and thus it must be applied \textit{on the convective side} of the boundary.

On the radiative side of the convective boundary, the situation is different since condition \ref{Sch} becomes
\begin{equation}
L_{rad} = L(m) \;\; \mathrm{and} \;\; \nabla_{rad} = \; \mathrm{or} \neq \nabla_{ad} \label{rad}
\end{equation}
depending on the presence, or not, of a discontinuity in chemical composition.

For a radiative layer to become convective, the Ledoux criterion$^1$ should be applied. For an equation of state $ P = \mathcal{R} \rho T / \beta \mu$ \rm with \rm $\beta = P_g/P$ and $\mu$, the mean molecular weight, it writes
\begin{equation}
\nabla_{rad} > \nabla_{ad} + \left ( \frac{\beta}{4 - 3 \beta} \right ) \frac{d\ln  \mu}{d\ln  P} = \nabla_{Ldx}  \label{Ldx}
\end{equation}

If the physical conditions within a layer are such that $\nabla_{rad}$ is in between $\nabla_{ad}$ and $\nabla_{Ldx}$, the layer is generally assumed to be semi-convective$^2$. In the absence of a special treatment of semi-convection in the code, such a layer will be assumed to be convective (radiative) if the Schwarzschild (Ledoux) criterion is adopted throughout the code. In MS massive stars for instance, the $\mu$-gradient region is characterized by a quasi-equality of $\nabla_{rad}$ and $\nabla_{ad}$. Small convective zones, leading to a step-like hydrogen profile, will generally appear with the Schwarzschild criterion while a smooth hydrogen profile is maintained throughout MS with the Ledoux criterion.

{\it Notes:\\
$^1$ Convection is the result of the non-linear development of the linear instability of dynamically unstable gravity modes, which is only the case if condition \ref{Ldx} is satisfied.\\
$^2$ Semi-convection is the result of the non-linear development of the linear vibrational instability of dynamically stable gravity modes.}

\subsection{Location of convective boundaries}
\label{boundary}
Whatever the convection criterion adopted in the code, the convective boundary is located by finding the zero of the function $y$, \textit{i.e.}
\begin{equation}
\begin{array}{lll}
&y &= \nabla_{rad} - \nabla_{ad} = 0 \;\;\;\; \mathrm{or} \label{bound} \\
&y &= \nabla_{rad} - \nabla_{Ldx} = 0
\end{array}
\end{equation}
\textit{on the convective side} of the boundary. However, a discontinuity in $y$ may be present at the boundary. With the Schwarzschild criterion, this happens when the chemical composition is discontinuous, \textit{i.e.} during the early MS phase of low mass stars ($\sim 1.1 -  1.6 M_{\odot}$) when the convective core is growing in mass, and during core helium-burning phase. With the Ledoux criterion, this happens all the time since, except in homogeneous models, the $\mu$-gradient is discontinuous at the boundary.

Finding the zero of a discontinuous function by interpolation or by locating a change of sign (or checking the sign layer by layer) leads to a infinity of solutions. Whatever the location of the discontinuity, \textit{i.e.} the location of the boundary, the condition on the change of sign will be satisfied. However, the only acceptable solution in the frame of LMLT is the one for which $\nabla_{rad} = \nabla_{ad}$ \textit{on the convective side} of the boundary. This indeed means that the radial convective velocity is zero and that the convective flux is accordingly zero at the boundary. This condition leads to $L_{rad} = L(m)$, which must be satisfied on both sides of the boundary. The way to proceed is to \textit{extrapolate} from points located exclusively in the convective region \citep[see the more complete discussion in][]{Gabriel14}.

The extent of convective cores of MS massive stars should be exactly identical, whatever the adopted convection criterion, if an extrapolation procedure is applied. With an interpolation (or change of sign) scheme however, the mass extent can be very different. If, at a given iteration, the estimated convective boundary is smaller than the real boundary, such a scheme will not be able to move the estimated boundary outward since the condition \ref{bound} will appear to be satisfied. This (too small) location will be accepted by the code as the correct one since from one iteration to the next, it will not move anymore.

\begin{figure}
\begin{center} 
\includegraphics[width=0.8\textwidth]{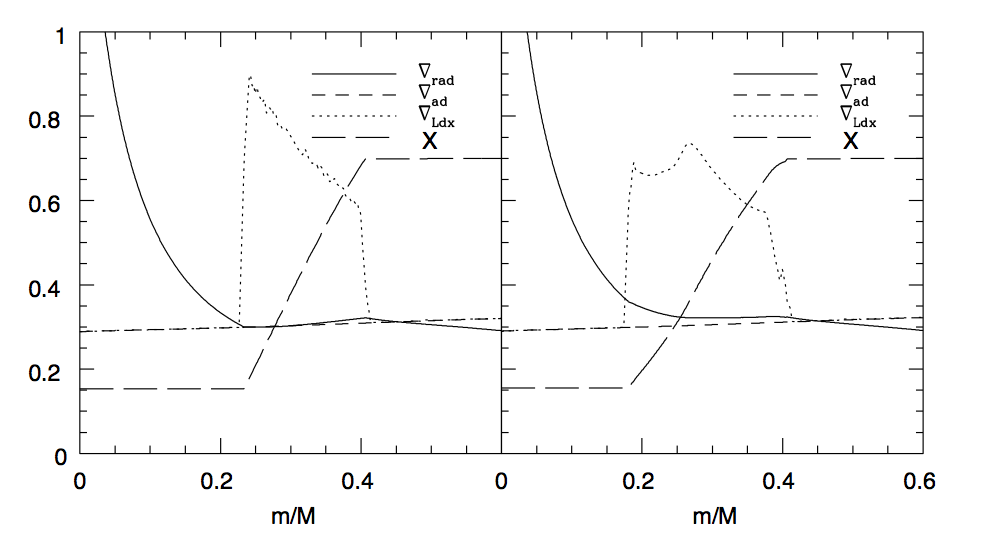} 
\caption{Hydrogen profile (long dashed line), radiative (full line), adiabatic (dashed line) and Ledoux (dotted line) temperature gradients, as a function of the fractional mass $m/M$, for an MS model of 16 $M_{\odot}$, computed with an extrapolation scheme (left panel) and computed with an interpolation scheme (right panel), using in both cases the Ledoux criterion}
\label{16M}
\end{center}
\end{figure}

Fig. \ref{16M} illustrates this behavior in the case of a $16 M_{\odot}$ MS star computed with the code CL\'ES \citep{Scuflaire08} and the Ledoux criterion. The hydrogen profile is drawn in long dashed line, $\nabla_{rad}$ in full line, $\nabla_{ad}$ in dashed line and $\nabla_{Ldx}$ in dotted line. In the left (right) panel, the location of the convective core boundary has been obtained through an extrapolation (interpolation) scheme.
One can easily see that the extent of the convective core (fully mixed region) is smaller when an interpolation scheme is used. At the end of MS, the difference can reach $\sim 25 \%$ depending on the distribution of mesh points for instance. It is also very clear that the model displayed in the right panel is not coherent since at the convective boundary, condition \ref{bound} is in fact not satisfied. Moreover, on top of the too small convective core, a misidentified semi-convective region covering the whole $\mu$-gradient region is found. This of course will lead to significantly different Brunt-V\"ais\"al\"a frequency distributions (see the behavior of $\nabla_{Ldx}$ in both panels).

\begin{figure}
\begin{center} 
\includegraphics[width=0.8\textwidth]{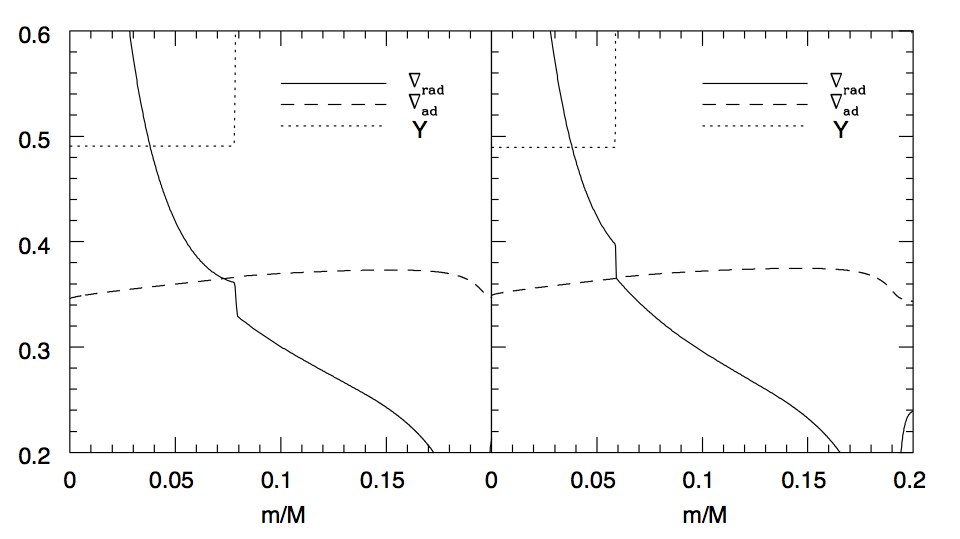} 
\caption{Helium profile (dotted line), radiative (full line), adiabatic (dashed line) temperature gradients, as a function of the fractional mass $m/M$, for a core helium-burning model of 8 $M_{\odot}$, computed with an extrapolation scheme (left panel) and computed with an interpolation scheme (right panel), using in both cases the Schwarzschild criterion}
\label{8M}
\end{center}
\end{figure}

Fig. \ref{8M} shows the helium profile (dotted line), $\nabla_{rad}$ (full line) and $\nabla_{ad}$ (dashed line) in a core helium-burning star of $8 M_{\odot}$ computed with the code CL\'ES and the Schwarzschild criterion. In the left (right) panel, the location of the convective core boundary results from an extrapolation (interpolation) scheme. 
The boundary of the convective core displayed in the left panel is fully consistent since it satisfies condition \ref{bound}.
A convective boundary similar to that displayed in the right panel of Fig. \ref{8M} was met and discussed by \citet{Castellani71}. With $\nabla_{rad}$ larger than $\nabla_{ad}$ at the boundary, $V_r$ is not equal to 0 and $L_{rad}$ is still smaller than $L(m)$ on the convective side of the boundary, while it must obviously be equal to $L(m)$ on the radiative side.
Indeed with the transformation of helium into carbon and oxygen, the discontinuity becomes larger and larger and \citet{Castellani71} showed that an increase of the convective core mass was the only way out of such an unstable situation. This is indeed obtained when an extrapolation procedure from convective points only is implemented in the code. 

It is clear that, with an underestimation of the extent of the convective core, the Brunt-V\"ais\"al\"a frequency distribution will be affected and as a result, the asteroseismic properties expected from core helium-burning stars will also be impacted.

\subsection{Convective shells}
\label{shell}
If the Ledoux criterion predicts a convective shell in a $\mu$-gradient region, the consistency of the model can be very difficult to achieve \citep[see Sect. 7 in][]{Gabriel14}. Fig. \ref{Shell} illustrates the change in the hydrogen profile arising from the occurrence of such a convective shell located in between mesh points $j_1$ and $j_2$. The opacity is noted $\kappa$ and the indices $i$ and $e$ refer to the inner and external sides of the convective boundaries. The main problems are listed below~:
\begin{itemize}
\item Two discontinuities arise except if the opacity, $\kappa$, is independent of the chemical composition. One of them will necessarily be such that $\kappa_e > \kappa_i$ and since $\nabla_{rad,i} = \nabla_{ad}$ on the inner side of both boundaries, $\nabla_{rad,e}$ will be greater than $\nabla_{ad}$ at the external side of either $j_1$ or $j_2$. This means that a semi-convective region may develop either below $j_1$ or above $j_2$.
\item Since consistent boundaries imply $L_{rad} = L(m)$ on both sides, $L_{rad}$ must necessarily decrease above $j_1$ and increase when reaching $j_2$. This might not be the case except maybe in nuclear burning shells or in the vicinity of an opacity peak. 
\item When consistency is not met, one of the boundaries is such that $\nabla_{rad} > \nabla_{ad}$ and $V_r \neq 0$. An overshooting or undershooting will take place and the chemical composition will change inside the shell, which will move along the $\mu$-gradient region.
\begin{figure}
\begin{center} 
\includegraphics[width=0.6\textwidth]{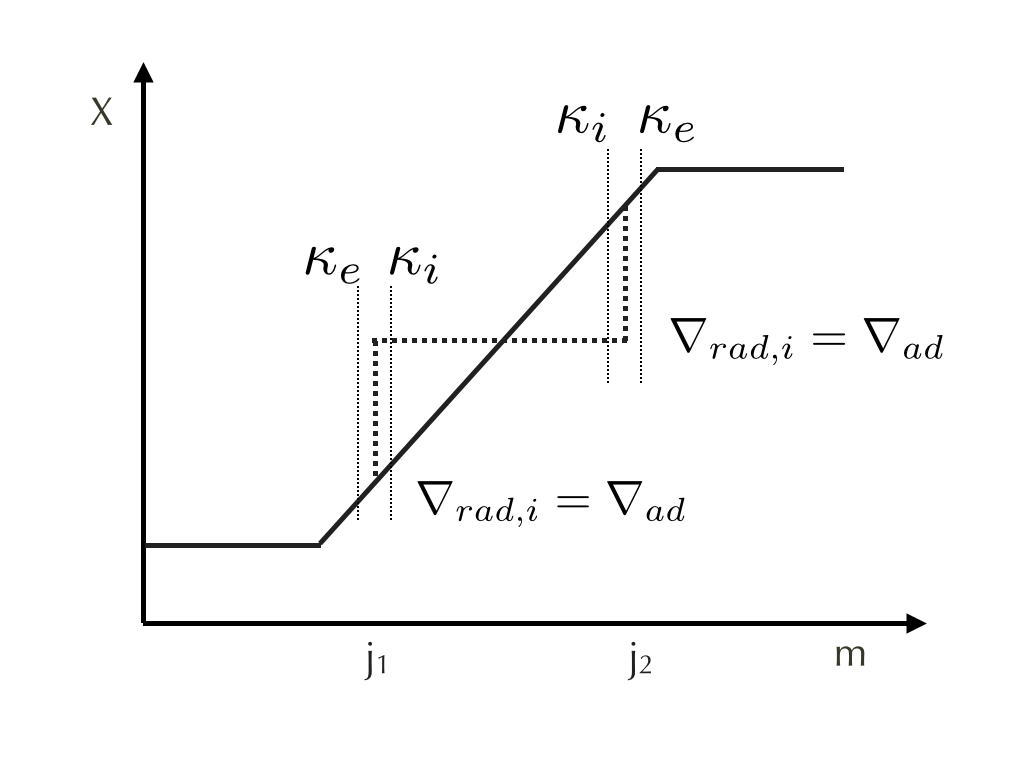} 
\caption{Schematic illustration of a convective shell developing  in a $\mu$-gradient region described by the hydrogen profile X as a function of m. $\kappa$ stands for the opacity coefficient and the indices $i$ and $e$ refer to the inner and external sides of the convective boundaries}
\label{Shell}
\end{center}
\end{figure}
\end{itemize}
Once more, the consistency should be checked at each convective boundary in order to correctly interpret the asteroseismic analyses.

\section{Intermediate convective zones (ICZ) in B supergiants}
\label{BSup} 
As can be seen in the left panel of Fig. \ref{16M}, MS models of massive stars are characterized by a near equality of $\nabla_{rad}$ and $\nabla_{ad}$ in the $\mu$-gradient region. Once the H-burning shell develops, the rapid increase of $L(m)/m$ creates a region where $\nabla_{Ldx} > \nabla_{rad} > \nabla_{ad}$, \textit{i.e.} a semi-convective zone, which is treated as convective if the Schwarzschild criterion is used. This intermediate convective zone (ICZ) overlaps the region of nuclear energy production. 

\begin{figure}[!h]
\includegraphics[width=\textwidth]{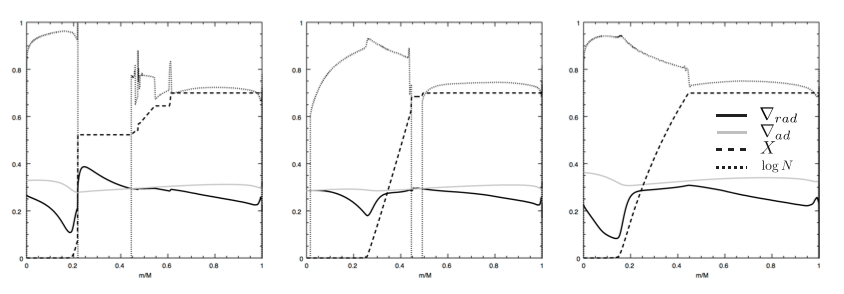} 
\caption{Hydrogen profile (dashed line), radiative (full line) and adiabatic (gray line) temperature gradients, and $\log{N}$ (dotted line) (N is the Brunt-V\"ais\"al\"a frequency) for a post-MS model of 16 $M_{\odot}$ still quite close to the MS turn-off, computed with the Schwarzschild criterion. Left panel~: no extra mixing, no mass loss. Middle panel~: extra-mixing with $\alpha_{em} = 0.2$. Right panel~: mass loss with $\dot{M} = 2 \; 10^{-7}M_{\odot}/yr$. The regions where N is equal to zero are convective}
\label{ICZSch}
\end{figure}

Fig. \ref{ICZSch} shows such an ICZ in a post-MS star of 16 $M_{\odot}$ computed with CL\'ES and the Schwarzschild criterion. The dashed line is the hydrogen profile, $\nabla_{rad}$ and $\nabla_{ad}$ are respectively drawn in black and gray full lines and the Brunt-V\"ais\"al\"a frequency (in $\log$) is shown in dotted line. 

The left panel illustrates a model computed without any extra-mixing nor any mass loss. If an extra-mixing is added on top of the convective core during MS, the near equality of both temperature gradients is replaced by $\nabla_{rad} < \nabla_{ad}$ in the $\mu$-gradient region and for large enough extra-mixing, the ICZ can become very small and be disconnected from the nuclear burning region. The middle panel of Fig.\ref{ICZSch} is an illustration of this influence of an extra-mixing ($\alpha_{em} = 0.2$). If the models are computed with mass loss, the near equality is also affected since the $\mu$-profile is less steep and this lessens the increase of the opacity, leading to smaller values of $\nabla_{rad}$. With a high enough mass loss rate, the ICZ does not appear, as can be seen in the right panel of Fig. \ref{ICZSch} ($\dot{M} =  2 \; 10^{-7} M_{\odot}/yr$). 

\begin{figure}[!h]
\includegraphics[width=\textwidth]{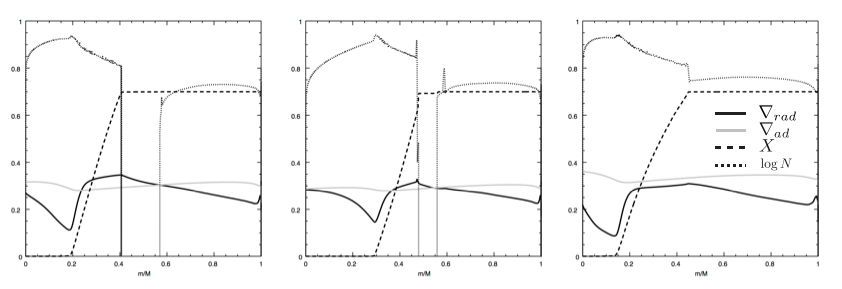} 
\caption{Hydrogen profile (dashed line), radiative (full line) and adiabatic (gray line) temperature gradients, and $\log{N}$ (dotted line) (N is the Brunt-V\"ais\"al\"a frequency) for a post-MS model of 16 $M_{\odot}$ still quite close to the MS turn-off, computed with the Ledoux criterion. Left panel~: no extra mixing, no mass loss. Middle panel~: extra-mixing with $\alpha_{em} = 0.3$. Right panel~: mass loss with $\dot{M} = 2 \; 10^{-7}M_{\odot}/yr$. The regions where N is equal to zero are convective}
\label{ICZLdx}
\end{figure}

When the Ledoux criterion is used, the semi-convective zone is treated as radiative and the ICZ is limited to the base of the homogeneous hydrogen-rich envelope. This is illustrated in the left panel of Fig. \ref{ICZLdx}. For the reasons discussed here above, the ICZ becomes smaller when an extra-mixing is added during MS (middle panel, $\alpha_{em} = 0.2$), and it disappears when a rather high mass loss rate is applied (right panel, $\dot{M} =  2 \; 10^{-7} M_{\odot}/yr$). With the Ledoux criterion, whatever the extent of the ICZ, its location is disconnected from the nuclear burning region.

The importance of an ICZ in post-MS stars has been emphasized by the discovery of excited g-modes in a B supergiant observed by the MOST satellite \citep{Saio06}. The role played by the ICZ is that of a barrier preventing g-modes from entering a very stabilizing helium core. The huge density in the central layers of such post-MS stars is indeed responsible for a strong radiative damping and in a star devoid of an ICZ, no g-modes can be excited \citep[see also][]{Gautschy09, Godart09}. The effects of extra-mixing and mass loss have been investigated and thoroughly discussed by \citet{Godart09} \citep[see also][and references therein]{Godart14}. 

Radial pulsations observed in $\alpha$ Cygni variables are also good indicators of the presence of an ICZ. As discussed in \citet{Saio13}, a high value of $L/M$ is required in order to obtain excited radial modes with periods compatible with those observed in $\alpha$ Cygni stars. Such high $L/M$ can only be met in massive stars coming back from the red supergiant (RSG) stage, after a heavy mass loss undergone as a RSG. Moreover for two stars, Rigel and Deneb, the superficial $N/C$ and $N/O$ ratios show nitrogen and oxygen enrichments typical of $CNO$ burning. With models computed with the Schwarzschild criterion, \citet{Saio13} obtained far too strong such enrichments, resulting from the overlap of the ICZ and the H-burning shell. More recently, \citet{Georgy14} computed similar models with the Ledoux criterion and showed that, with the disconnection of the ICZ and the H-burning shell, the enrichments were much closer to the observations. 

The \textit{power of asteroseismology} is here clearly seen in those two examples~:
\begin{itemize}
\item One single g-mode observed in a B supergiant star is a signature of an ICZ and therefore brings constraints on the amount of extra-mixing and mass loss.
\item  The detection of radial modes, coupled with a detailed spectroscopic analysis, in an $\alpha$ Cygni variable star, not only lifts the degeneracy between supergiant models crossing the Hertzsprung-Russell diagram from blue to red and vice-versa but also imposes constraints on the location and mass extent of the ICZ.
\end{itemize}

\section{Opacity in the metal opacity bump}
\label{opacity}
Since the excitation mechanism in $\beta$ Cephei  and SPB stars is the $\kappa$-mechanism acting in the metal opacity bump at $T \simeq 2 \; 10^5 K$, it is no surprise that metallicity plays a keyrole in the extension of their instability strips in the Hertzsprung-Russell diagram. It was indeed shown by \citet{Miglio07} that at $Z = 0.005$ the instability strip for $\beta$ Cephei stars does no longer exist while a very narrow strip still remains for SPB stars. Those results actually predict that no $\beta$ Cephei and very few SPB stars should be expected in the Magellanic Clouds. 

Challenging observations of such stars in the Small Magellanic Cloud (SMC) have however been presented by \citet{Karoff08, Diago08, Kourniotis14}. The possible explanations for this apparent contradiction have been investigated by \citet{Salmon12} who concluded their analysis by suggesting that \textit{the current opacity data are underestimating the stellar opacity due to nickel by a factor $\sim 2$ (in the metal opacity bump)}. This has indeed been an additional incentive element to proceed to new revisions of the stellar opacities. Recent comparisons between different opacity codes presented by \citet{Turck13} are very encouraging since they evoke a possible increase of the nickel opacity in the metal opacity bump by a factor similar to that proposed by \citet{Salmon12}.

This is not the first time that variable stars have demanded a revision in opacity data. \citet{Simon82}'s plea was followed by a successful update, which turned out to be the key factor in explaining the excitation mechanism in $\beta$ Cephei stars. As for now, the OP, OPAC and OPAL teams are undertaking new opacity computations, especially in physical conditions typical of B-type and solar-like stars. 

\section{Conclusions}
\label{Conclusions}
Asteroseismology is definitely a powerful tool to bring some light on some unsolved problems affecting the structure and evolution of massive stars. However, we have tried with a few selected examples to call attention on some aspects of stellar modeling as well as asteroseismic analyses, which could lead to misinterpretations of seismic data~:

\begin{itemize}
\item Our \textit{first word of caution} relates to the asteroseismic estimation of the extent of extra-mixing on top of the convective core in $\beta$ Cephei stars. Although not sufficient the following requirements must be met in order to have a really reliable determination~:
\begin{itemize}
\item at least 4-5 \textit{well identified} frequencies must be observed;
\item the metallicity, and more precisely the detailed elemental abundances, should be known.
\end{itemize}
\item Our \textit{second word of caution} sets in the LMLT frame of convection. For a theoretically asteroseismic analysis to be reliable, fully consistent models are absolutely required. This implies that~:
\begin{itemize}
\item the condition on convective neutrality defining the boundary of a convective region must be applied on the \textit{convective side} of the boundary;
\item when this boundary is affected by a discontinuity in the chemical composition (Schwarzschild criterion) or in the $\mu$-gradient (Ledoux criterion), the location of the boundary must result from an \textit{extrapolation} from convective points only;
\item the boundaries of each intermediate convective zones should systematically be checked for consistency.
\end{itemize}
\item Our \textit{third word of caution} addresses the physical data entering stellar model computations. It certainly is another success of asteroseismology to have been at the origin of a new, and still in progress, opacity revision. This stresses the necessity to be open-minded to all the tools available in order to unveil the physical processes inside massive stars, \textit{i.e.} not only asteroseismology but also spectroscopy and photometry for detailed elemental abundances and global stellar properties, as well as the most updated physical data related to nuclear reactions and opacities.

\end{itemize}

\bibliographystyle{iau307}
\bibliography{Noels}

\begin{thebibliography}{}

\bibitem[\protect\astroncite{{Aerts}}{2014}]{Aerts14}
{Aerts}, C. 2014,
\newblock in G. {Meynet}, C. {Georgy}, J.~H. {Groh}, \& P. {Stee} (eds.), {\em
  IAU Symposium}, Vol. 307 of {\em IAU Symposium}

\bibitem[\protect\astroncite{{Asplund} et~al.}{2005}]{Asplund05}
{Asplund}, M., {Grevesse}, N., \& {Sauval}, A.~J. 2005,
\newblock in T.~G. {Barnes}, III \& F.~N. {Bash} (eds.), {\em Cosmic Abundances
  as Records of Stellar Evolution and Nucleosynthesis}, Vol. 336 of {\em
  Astronomical Society of the Pacific Conference Series}, p.~25

\bibitem[\protect\astroncite{{Badnell} et~al.}{2005}]{Badnell05}
{Badnell}, N.~R., {Bautista}, M.~A., {Butler}, K., {et~al.} 2005,
\newblock {\em \mnras} 360, 458

\bibitem[\protect\astroncite{{Castellani} et~al.}{1971}]{Castellani71}
{Castellani}, V., {Giannone}, P., \& {Renzini}, A. 1971,
\newblock {\em \apss} 10, 340

\bibitem[\protect\astroncite{{Diago} et~al.}{2008}]{Diago08}
{Diago}, P.~D., {Guti{\'e}rrez-Soto}, J., {Fabregat}, J., \& {Martayan}, C.
  2008,
\newblock {\em \aap} 480, 179

\bibitem[\protect\astroncite{{Gabriel} et~al.}{2014}]{Gabriel14}
{Gabriel}, M., {Noels}, A., {Montalban}, J., \& {Miglio}, A. 2014,
\newblock {\em ArXiv e-prints}

\bibitem[\protect\astroncite{{Gautschy}}{2009}]{Gautschy09}
{Gautschy}, A. 2009,
\newblock {\em \aap} 498, 273

\bibitem[\protect\astroncite{{Georgy} et~al.}{2014}]{Georgy14}
{Georgy}, C., {Saio}, H., \& {Meynet}, G. 2014,
\newblock {\em \mnras} 439, L6

\bibitem[\protect\astroncite{{Godart} et~al.}{2014}]{Godart14}
{Godart}, M., {Grotsch-Noels}, A., \& {Dupret}, M.-A. 2014,
\newblock in J.~A. {Guzik}, W.~J. {Chaplin}, G. {Handler}, \& A. {Pigulski}
  (eds.), {\em IAU Symposium}, Vol. 301 of {\em IAU Symposium}, pp 313--320

\bibitem[\protect\astroncite{{Godart} et~al.}{2009}]{Godart09}
{Godart}, M., {Noels}, A., {Dupret}, M.-A., \& {Lebreton}, Y. 2009,
\newblock {\em \mnras} 396, 1833

\bibitem[\protect\astroncite{{Grevesse} \& {Noels}}{1993}]{Grevesse93}
{Grevesse}, N. \& {Noels}, A. 1993,
\newblock in B. {Hauck}, S. {Paltani}, \& D. {Raboud} (eds.), {\em
  Perfectionnement de l'Association Vaudoise des Chercheurs en Physique}, pp
  205--257

\bibitem[\protect\astroncite{{Iglesias} \& {Rogers}}{1996}]{Iglesias96}
{Iglesias}, C.~A. \& {Rogers}, F.~J. 1996,
\newblock {\em \apj} 464, 943

\bibitem[\protect\astroncite{{Karoff} et~al.}{2008}]{Karoff08}
{Karoff}, C., {Arentoft}, T., {Glowienka}, L., {et~al.} 2008,
\newblock {\em \mnras} 386, 1085

\bibitem[\protect\astroncite{{Kourniotis} et~al.}{2014}]{Kourniotis14}
{Kourniotis}, M., {Bonanos}, A.~Z., {Soszy{\'n}ski}, I., {et~al.} 2014,
\newblock {\em \aap} 562, A125

\bibitem[\protect\astroncite{{Miglio} et~al.}{2007}]{Miglio07}
{Miglio}, A., {Montalb{\'a}n}, J., \& {Dupret}, M.-A. 2007,
\newblock {\em \mnras} 375, L21

\bibitem[\protect\astroncite{{Paxton} et~al.}{2013}]{Paxton13}
{Paxton}, B., {Cantiello}, M., {Arras}, P., {et~al.} 2013,
\newblock {\em \apjs} 208, 4

\bibitem[\protect\astroncite{{Saio} et~al.}{2013}]{Saio13}
{Saio}, H., {Georgy}, C., \& {Meynet}, G. 2013,
\newblock {\em \mnras} 433, 1246

\bibitem[\protect\astroncite{{Saio} et~al.}{2006}]{Saio06}
{Saio}, H., {Kuschnig}, R., {Gautschy}, A., {et~al.} 2006,
\newblock {\em \apj} 650, 1111

\bibitem[\protect\astroncite{Salmon}{2014}]{Salmon14}
Salmon, S. 2014,
\newblock {\em Ph.D. thesis}, University of Li\`ege, Belgium

\bibitem[\protect\astroncite{{Salmon} et~al.}{2012}]{Salmon12}
{Salmon}, S., {Montalb{\'a}n}, J., {Morel}, T., {et~al.} 2012,
\newblock {\em \mnras} 422, 3460

\bibitem[\protect\astroncite{{Scuflaire} et~al.}{2008}]{Scuflaire08}
{Scuflaire}, R., {Th{\'e}ado}, S., {Montalb{\'a}n}, J., {et~al.} 2008,
\newblock {\em \apss} 316, 83

\bibitem[\protect\astroncite{{Simon}}{1982}]{Simon82}
{Simon}, N.~R. 1982,
\newblock {\em \apjl} 260, L87

\bibitem[\protect\astroncite{{Turck-Chi{\`e}ze} \& {Gilles}}{2013}]{Turck13}
{Turck-Chi{\`e}ze}, S. \& {Gilles}, D. 2013,
\newblock in {\em European Physical Journal Web of Conferences}, Vol.~43 of
  {\em European Physical Journal Web of Conferences}, p. 1003

\end{thebibliography}

\begin{discussion}

\discuss{C. Aerts}{Just a general remark for the majority of the audience about the hare-and-hound exercises~: you should also compare the uncertainty on the model parameters (M, X, Z, $\alpha_{em}$) between the case where you have only $T_{eff}$ and $\log g$, with errors of 1000K and 0.2 dex, and the case where you also have a few identified modes, including, or not, a misidentification and/or wrong input physics.}

\discuss{A. Noels}{I fully agree that most of the time, the addition of 4 or 5 well identified modes will highly increase the level of constraints on the theoretical model best fitting the observed star. This is indeed the power of asteroseismology to probe deep inside the star, down to the centrally mixed region. This lifts the well-known degeneracy of possible $M$ and $\alpha_{em}$ for a given set of $T_{eff}$ and $L$. However, I want to call attention on a possible misinterpretation of asteroseismic data if the physics is not the same in the grid and in the observed target. It is of course impossible to check all the possible differences but an extra care should be given to the detailed chemical composition. This is indeed, in my opinion, an \textit{important word of caution}~: to increase the relevance and the power of asteroseismic analyses, all the tools should enter the game, and among them, detailed spectroscopic and photometric analyses definitely play key roles.}

\end{discussion}

\end{document}